\documentclass[11pt]{article}
\usepackage{amsmath, amsfonts, amssymb}
\usepackage{graphicx}

\begin{document}
\date{}

\title{Single pairs of time-bin entangled photons}

\author{
Marijn A. M. Versteegh$^{1,2,3}$, Michael E. Reimer$^{1,4}$,\\ Aafke A. van den Berg$^{1}$,
 Gediminas Juska$^{5}$,\\ Valeria Dimastrodonato$^{5}$, Agnieszka Gocalinska$^{5}$,\\
  Emanuele Pelucchi$^{5}$, Val Zwiller$^{1}$\\
\\
\small $^{1}$Kavli Institute of Nanoscience, Delft University of Technology,\\
\small Lorentzweg 1, 2628 CJ Delft, The Netherlands\\
\small $^{2}$Quantum Optics, Quantum Nanophysics and Quantum Information\\
\small Faculty of Physics, University of Vienna,\\
\small Boltzmanngasse 5, 1090 Vienna, Austria\\
\small $^{3}$Institute for Quantum Optics and Quantum Information,\\
\small Austrian Academy of Sciences, Boltzmanngasse 3, 1090 Vienna, Austria\\
\small $^{4}$Institute for Quantum Computing and Department of Electrical \& Computer\\
\small Engineering, University of Waterloo, Waterloo, Ontario N2L 3G1, Canada\\
\small $^{5}$Tyndall National Institute, University College Cork,\\
\small Lee Maltings, Cork, Ireland} \maketitle

\begin{abstract}
\noindent
Time-bin entangled photons are ideal for long-distance quantum communication via optical fibers. Here we present a source where, even at high creation rates, each excitation pulse generates at most one time-bin entangled pair. This is important for the accuracy and security of quantum communication. Our site-controlled quantum dot generates single polarization-entangled photon pairs, which are then converted, without loss of entanglement strength, into single time-bin entangled photon pairs.
\end{abstract}

\section{INTRODUCTION}
Entanglement of flying qubits is a fundamental principle of quantum information and communication \cite{gisin2007, pan2012}, and is at the basis of quantum communication protocols such as quantum key distribution \cite{ekert1991}, quantum teleportation \cite{bennett1993}, and quantum secure direct communication \cite{deng2003}.  Commonly used qubits are polarization-entangled photons generated by spontaneous parametric down-conversion \cite{kwiat1995}. There are three important limitations to this approach.

First, in optical fibers, polarization encoding is prone to thermal and mechanical disturbances, which affect the fiber's birefringence and thereby also the polarization of the outcoupled photons. This phenomenon, called polarization mode dispersion, is a major problem for quantum communication in real-world implementations \cite{brodsky2011, antonelli2011}. The solution is time-bin entanglement \cite{franson1989, brendel1999}: quantum information encoded in the arrival time of photons. Time-bin entanglement is robust \cite{thew2002, marcikic2004} and has enabled distribution of entangled photons over 300 km of optical fiber \cite{inagaki2013}.

Second, parametric down-conversion is a random process and follows Poissonian statistics. If $p$ is the probability for a pump pulse to create a pair of entangled photons, there is a probability of order $p^2$ to create two or more pairs of entangled photons from the same pump pulse. Generation of multiple pairs makes it unclear which photon is entangled with which, and thus reduces the accuracy and security of the quantum communication. A possible solution is to reduce the pair-creation rate, so that $p^2$ becomes very small. Of course, the data transmission rate is then reduced as well. A more rigorous solution, which has been shown to work for polarization entanglement (but not yet for time-bin entanglement), is to generate single pairs of entangled photons by exciting a biexciton (XX) in a semiconductor quantum dot \cite{benson2000, akopian2006, young2006, hafenbrak2007, stevenson2008, muller2009, salter2010, dousse2010, juska2013}. A XX is an excited state consisting of two electrons and two holes. Recombination of one electron and one hole occurs under emission of one photon (the XX photon), and brings the quantum dot to the exciton (X) state, which then decays further to the ground state, again emitting one photon (the X photon). If the excitation laser pulse is longer than the lifetime of the XX (typically around 1 ns), re-excitation and multiple-pair emission can occur. In contrast, if the excitation laser pulse is much shorter than the XX lifetime, each laser pulse can create not more than one photon pair.

Third, there is the problem of scalability. A great challenge is to make large arrays of identical entangled photon sources, for use in quantum information applications. Quantum dots offer scalable, nano-sized sources of entangled photons. Quantum dot sources are energy efficient, in the sense that much less optical pump power is required per entangled photon pair as with downconversion sources. Some of us recently reported an array of site-controlled pyramidal InGaAs$_{1-\delta}$N$_\delta$ quantum dots, with areas where up to 15\% of the quantum dots emit polarization entangled photons \cite{juska2013, juska2015}.

Is it possible to combine the afore-mentioned solutions, and generate single pairs of time-bin entangled photons? In 2005 Simon and Poizat suggested a method \cite{simon2005}, but the lack of a suitable metastable state in the quantum dot, essential in their proposal, has prevented implementation. A recent approach by Jayakumar \textit{et al.} \cite{jayakumar2014} is based on a quantum dot, but still allows for a probability of order $p^2$ to create two photon pairs from the same incoming laser pulse, because they split their pulse into two.

Here, we demonstrate the first source of single time-bin entangled photon pairs, using a site-controlled quantum dot. We overcome the need of a metastable state and avoid the multi-photon emission probability $p^2$ that is inherent to all previous implementations of time-bin entanglement generation. Our source opens up new possibilities for transfer of spin-photon entanglement \cite{togan2010, degreve2012, gao2012, pfaff2014} over long distances, hyperentanglement \cite{barreiro2005}, quantum dense coding \cite{schuck2006} and deterministic entanglement purification \cite{sheng2014}, and could be developed further for integration in compact and scalable quantum information devices.

\section{POLARIZATION-TIME-BIN INTERFACE}

In our experiment, a site-controlled semiconductor quantum dot generates single photon pairs in the polarization entangled state $(|HH\rangle+e^{i\varphi}|VV\rangle)/\surd2$, where $H$ ($V$) stands for horizontal (vertical) polarization and $\phi$ is a phase angle. The probability of multiple pair emission from one excitation pulse is strongly suppressed. The polarization entangled state is then converted into the time-bin-entangled state $(|ee\rangle+e^{i(\phi+\sigma)}|ll\rangle)/\surd2$, where $e$ ($l$) stands for the early (late) time bin. This conversion occurs in a polarization-time-bin interface, consisting of an unbalanced Mach-Zehnder interferometer with polarizing beamsplitters (PBSs) and a polarizer (Fig. 1). The horizontal (vertical) term of the two-photon wave function takes the short (long) path, and is thus converted into the early (late) term of the time-bin entangled wave function. The path length difference determines the separation between the two time bins, 4.3 ns in our case, and also the phase $\sigma$. The polarizer at $45^{\circ}$ behind the second PBS erases all polarization entanglement, leaving only the time-bin entanglement. In a quantum communication network, where messages are encoded in the polarization bases, this interface can realize the transfer of quantum information to the time-bin bases. The resulting time-bin entangled qubits could be faithfully transmitted through an optical fiber suffering from mechanical vibrations or thermal instability. The 50\% intensity loss at the polarizer could in principle be avoided by operating a fast Pockels cell in front of the polarizer. This Pockels cell should then switch between the early time bin and the late time bin, and, by polarization rotation, ensure that the early and late parts of the wave function obtain the same polarization. Other realizations of conversion between polarization and time-bin entanglement (with spontaneous parametric down-conversion) are described in Refs. \cite{sanaka2002, takesue2005, bussieres2010, martin2013}.

\section{XX-X RADIATIVE CASCADE}
\subsection{Pyramidal quantum dot}

In the present experiment, the single pairs of polarization entangled photons are produced in one pyramidal quantum dot \cite{dimastrodonato2012}. Our sample consists of an array of pyramids [Fig. 2(a)], where each pyramid contains a single embedded quantum dot. Figure 2(b) shows a schematic sketch of the internal structure of a pyramid, consisting of several epitaxial layers. The pyramids were grown by metal-organic vapor phase epitaxy in 7.5-$\mu$m-pitch tetrahedral recesses etched in (111)B-oriented GaAs. Details of the growth method are described in Ref. \cite{juska2013}.

\subsection{Excitation}
We excited one pyramidal quantum dot with 639-nm 100-ps 80-MHz laser pulses. The diameter of the excitation spot was 1 $\mu$m, much smaller than the 7.5-$\mu$m distance between the pyramids, so that we could easily capture the optical emission from just one quantum dot. The emission spectrum [Fig. 2(c)] shows clear XX, X, and trion (X*) emission lines. In our experiment, we spectrally selected the XX and X emission. Weaker emission lines on the sides [visible in Fig. 2(c)] appear only under strong excitation of the quantum dot, when a significant fraction of the pulses creates more than two electron-hole pairs in the quantum dot. In such case, charge carriers at higher energy levels, by electrostatic interaction, shift the emission energy of the charges at the lowest energy levels. By spectrally rejecting the side emission lines we made sure that for one excitation pulse only one XX photon and one X photon were measured. During all measurements the sample was maintained at a constant temperature of 5.1 K in a closed-cycle cryostat. Except when stated otherwise, we used an excitation power of 150 nW at the sample. At this power the average number of created electron-hole pairs per excitation pulse was 0.5.

\subsection{Characterization of the XX-X radiative cascade}

We observed that the XX line has a quadratic dependence on power, the X line a nearly linear dependence up to saturation [Fig. 3(a)]. Time-resolved photoluminescence measurements show a XX lifetime of 0.72 $\pm$ 0.03 ns and an X lifetime of 1.25 $\pm$ 0.04 ns [Fig. 3(b)]. Both XX and X emission is unpolarized [Fig. 3(c,d)]. We measured a fine-structure splitting of $S=0.6\pm0.2$ $\mu$eV [Fig. 3(e)], corresponding to a precession period of the X spin state of $h/S=7.3\pm1.9$ ns, where $h$ is Planck's constant \cite{stevenson2008}. Since this period is much longer than the X lifetime, spin precession has only a small influence on the correlations between the polarization of the XX and X photons. This small value for the fine-structure splitting, which is a special feature of our pyramidal quantum dots \cite{juska2013}, enables measurement of quantum entanglement without the need for strict temporal post-selection and is thus of great importance for practical implementations of quantum communication.

\section{SINGLE PAIRS OF POLARIZATION-ENTANGLED PHOTONS}
\subsection{Single-photon correlation measurements}

Single-photon time-resolved correlation measurements were performed by splitting the emission into two arms, each containing a spectrometer and an avalanche photodiode (APD). The outputs of the two APDs were connected to the time tagging module, which registers the differences in arrival time between signals from both APDs. Hanbury-Brown Twiss autocorrelation measurements, to determine the sub-Poissonian statistics, were performed by selecting with both spectrometers the X emission, or selecting with both spectrometers the XX emission. Time-resolved cross-correlation measurements, to study the emission from the XX-X radiative cascade, were performed by selecting with one spectrometer the XX emission and with the other the X emission.

\subsection{Single-photon statistics}
Hanbury-Brown Twiss autocorrelation measurements gave $g^{(2)}(0)=0.13$ as maximum for the central peak for X photons [Fig. 4(a)], and $g^{(2)}(0)=0.22$ for XX photons [Fig. 4(b)], while XX-X cross correlation measurements gave $g^{(2)}(0)=3.5$ [Fig. 4(c)]. These correlation results show that the quantum dot is a sub-Poissonian source, i.e., a single-photon source, of photon pairs from the XX-X cascade. Note that for a Poissonian source, a downconversion source for example, the central peak in an autocorrelation measurement is as high as the neighboring peaks.

The measured nonzero value for $g^{(2)}(0)$ in the autocorrelation measurements can be explained by re-excitation. Some biexcitons and excitons decay within the 100 ps long excitation pulse, so that the quantum dot can be excited again by a pump photon, leading to emission of more than one XX photon or X photon. Purer single-photon statistics can thus be obtained by reducing the excitation pulse duration. We therefore also took Hanbury-Brown Twiss autocorrelation measurements with 750-nm 3-ps 80-MHz excitation laser pulses at the same power of 150 nW. We found $g^{(2)}(0)=0.03$ as maximum for the central peak for X photons [Fig. 4(d)], and $g^{(2)}(0)=0.05$ for XX photons (integration over the peak gives $g^{(2)}=0.07$) [Fig. 4(e)], while XX-X cross correlation measurements under the same excitation conditions gave $g^{(2)}(0)=4.6$ [Fig. 4(f)]. These correlation results show that the quantum dot generates nearly perfect single photon pairs from the XX-X cascade, emitting not more than one X photon and not more than one XX photon for each excitation pulse. For all other measurements we used our 100-ps laser, because its power was more stable.

\subsection{Polarization entanglement}

XX-X cross-correlation measurements with polarization selection in the rectilinear, diagonal and circular bases show that the two photons from the XX-X cascade are polarization entangled [Fig. 5(a)]. This can be seen from the fact that $HH$ is more than twice as strong as $HV$, $DD$ is more than twice as strong as $DA$, and $RL$ is more than twice as strong as $LL$. Here, the first letter represents the polarization of the XX photon, the second letter the polarization of the X photon. $D$ ($A$) is diagonal (antidiagonal) polarization and $L$ ($R$) is left-handed (right-handed) circular polarization.

In order to determine the full quantum state of the photon pair, a standard quantum state tomography was performed, following the method described by James \textit{et al.} \cite{james2001}. We performed 16 time-resolved cross-correlation measurements, with one spectrometer selecting the XX emission and the other the X emission, with the following polarization selections: $HH$, $HV$, $HD$, $HL$, $VH$, $VV$, $VD$, $VL$, $DH$, $DV$, $DD$, $DL$, $LH$, $LV$, $LD$, and $LL$.  The polarization selections were made using quarter wave plates, half wave plates and polarizers. Additional half wave plates were used to ensure that always the same polarization enters the spectrometers, and thus to avoid any effect of polarization sensitivity of the spectrometers on the results. For each measurement we used an integration time of 300 s.

The observed density matrix [Fig. 5(b,c)] shows a polarization entangled state with a concurrence of $0.54\pm0.03$, where a positive value indicates quantum entanglement \cite{wootters2001}. The measured coincidence numbers, on which this density matrix is based, are given in as Supplemental Material \cite{supplement}. The fidelity to the maximally entangled state $(|HH\rangle+|VV\rangle)/\surd2$ is $0.722\pm0.006$, where 0.5 is the classical limit. The fidelity to $(|HH\rangle+e^{0.141\pi i}|VV\rangle)/\surd2$, also a maximally entangled state, is $0.758\pm0.006$. It has been observed before that the highest fidelity in quantum dots is found with respect to a state $(|HH\rangle+e^{i\varphi}|VV\rangle)/\surd2$ with a small phase $\varphi$ \cite{dousse2010}. This phase could be attributed to precession of the X spin state, or to a birefringence in the sample or in the setup. In our calculations we used a time window of 3 ns, which is 2.4 times the X lifetime, thus including more than 90\% of the correlation counts. For the calculation of the density matrix we used a maximum likelihood estimation, following Ref. \cite{james2001}. The fidelity $F(\rho,\psi)$ of a density matrix $\rho$ to a pure state $|\psi\rangle$ is calculated from $F=\langle \psi|\rho|\psi\rangle$.

\section{SINGLE PAIRS OF TIME-BIN ENTANGLED PHOTONS}
\subsection{Time-bin quantum state tomography}

To obtain single time-bin entangled photon pairs, we converted the single polarization entangled photon pairs from the pyramidal quantum dot using our polarization-time-bin interface (Fig. 1). As already mentioned, the polarizer in this interface eliminates all polarization entanglement. We analyzed the time-bin quantum state by time-bin quantum state tomography \cite{bussieres2010, takesue2009}. The photons have to be sent again through an unbalanced interferometer, so that the early and late terms of the two-photon wave function overlap with each other. Essentially, we used the method of Bussi\`{e}res \textit{et al.} \cite{bussieres2010}, where PBSs in the interferometer enable the time-bin quantum state tomography by employing waveplates and polarizers in a similar way as in polarization quantum state tomography. For the time-bin quantum state tomography we used an integration time of 1800 s for each measurement.

The essence of this form of time-bin quantum state tomography is that the time-bin entanglement is converted back into polarization entanglement. In our experiment, this conversion back into polarization entanglement was established by two elements in our setup. First, the time-bin entangled photons were sent back through the unbalanced Mach-Zehnder interferometer, as indicated by the orange paths in Fig. 1. Second, we measured with our correlation electronics not just the time difference between the XX photon and the X photon, as in the regular polarization quantum state tomography, but we measured the timing of the XX photon and the X photon with respect to the trigger signal from our pulsed excitation laser. Based on their arrival times with respect to the laser pulse, the detected photons were discriminated into three categories: photons that traveled twice the short path in the interferometer, photons that traveled once the short path and once the long path (here it is fundamentally uncertain whether they first traveled the short path and then the long path or first the long path and then the short path), and photons that traveled twice the long path [Fig. 6(a)]. The photon pairs traveling twice the short (long) path were with certainty in the $|ee\rangle$ ($|ll\rangle$) state and are therefore, as a result of our measurement, not time-bin entangled. The pairs of photons that traveled once the short path and once the long path were entangled and therefore we temporally post-selected those photons, again using a time window of 3 ns. We measured the correlations in the polarizations of the post-selected pairs. Here, a measured $V$ ($H$) polarization means that the photon was in the early (late) state. Likewise, a measured $D$ polarization corresponds to $(|l\rangle+|e\rangle)/\surd2$, $A$ corresponds to $(|l\rangle-|e\rangle)/\surd2$, $L$ to $(|l\rangle-i|e\rangle)/\surd2$, and $R$ to $(|l\rangle+i|e\rangle)/\surd2$. Thus, time-bin entanglement can be measured by measuring polarization entanglement. The post-selection, and the concomitant loss of intensity, could in principle be avoided by polarization rotation with a fast Pockels cell.

\subsection{Results}
The resulting density matrix is shown in Figs. 6(b) and 6(c). The measured coincidence numbers, on which this density matrix is based, are given as Supplemental Material \cite{supplement}. The concurrence is $0.58\pm0.07$, which demonstrates time-bin entanglement. The fidelity to $(|ee\rangle+e^{0.672\pi i}|ll\rangle)/\surd2$ is $0.74\pm0.02$. These values are, within experimental error, the same as the values we had obtained for polarization entanglement under the same excitation conditions, showing that the conversion takes place without loss of entanglement strength. As a result of traveling twice through the interferometer, the phase $\chi$ between the two components of the wave function is different from the phase that was measured with the polarization quantum state tomography.

\section{DISCUSSION}
Compared to Poissonian parametric-downconversion sources of time-bin entangled photons, our pyramidal quantum dot source has several advantages. First, the relative rate of pulses where more than one pair is emitted is strongly reduced, as shown by the antibunching data [Fig. 4]. Second, our type of sample offers the potential of scalability. The pyramidal quantum dot used in this study is just one among tens of thousands very similar position-controlled quantum dots on the sample, as indeed, site-controlled pyramidal quantum dots have demonstrated to be the highest uniformity quantum-dot system to date \cite{leifer2007, mereni2009}. Current research efforts in quantum dot growth will result in fabrication techniques with even higher control on shapes, composition and uniformity. The goal is to make arrays where all quantum dots emit strongly entangled photons in nearly identical quantum states. Third, the generation efficiency (in terms of power) is much higher than for non-linear crystals.

It must be said that downconversion sources still reach better entanglement fidelities. Yet, quantum dot sources show rapid progress towards purer (polarization-) entangled states \cite{young2009, kuroda2013, muller2014, trotta2014, versteegh2014}. A new approach is to create polarization entangled photons in a [111] grown quantum dot embedded in a nanowire \cite{versteegh2014, huber2014}, where the cylindric symmetry reduces the fine-structure splitting \cite{singh2009}. The nanowire shape additionally offers high brightness \cite{reimer2012} and coherent emission \cite{versteegh2014}, as well as a directional Gaussian emission profile \cite{bulgarini2014}.

Possible applications of single pairs of time-bin entangled photons include quantum communication via fiber, where, thanks to the time-bin encoding, the two-photon quantum state is not affected by fiber instabilities. A grating could be used to split the XX beam from the X beam. After transmission through the fibers it can be useful to convert the time-bin entanglement back into polarization entanglement, as is done in our setup, because the photon polarization can be easily manipulated. Other possible applications involve hyperentanglement (simultaneous polarization and time-bin entanglement), which could be obtained by removing the polarizer behind the interferometer in Fig. 1. Schuck \textit{et al.} \cite{schuck2006} have shown that it is possible to use time-bin entanglement, in addition to polarization entanglement, for a more complete Bell state analysis and quantum dense coding. Finally, by expanding the setup to a form like the one proposed by Sheng and Zhou \cite{sheng2014}, it should be possible to realize deterministic polarization entanglement purification with single photon pairs.

\section{CONCLUSION}
We generated single time-bin entangled photon pairs from a pyramidal quantum dot. The strength of entanglement is maintained in our polarization-time-bin interface. Our source of single time-bin entangled photons could be used in a quantum communication scheme, where entanglement is preserved in optical fibers suffering from mechanical or thermal vibrations. If desired, the entanglement can be converted back into the polarization bases, so that the photons could be further processed by widely available polarization sensitive optical components. Recently, generation of entangled photons from a semiconductor quantum dot at a telecom wavelength has been realized \cite{ward2014}. By combination with our technique, and with resonant excitation \cite{jayakumar2014, muller2014} and electrical injection \cite{salter2010}, one could make an electrically driven on-demand source of single time-bin entangled photons at a telecom wavelength.

\section{ACKNOWLEDGEMENTS}
This work was supported by the Dutch Foundation for Fundamental Research on Matter (FOM projectruimte 12PR2994), ERC, and the European Union Seventh Framework Programme 209 (FP7/2007-2013) under Grant Agreement No. 601126 210 (HANAS), the Irish Higher Education Authority Program for Research in Third Level Institutions (2007-2011) via the INSPIRE programme, by Science Foundation Ireland (grants 05/IN.1/I25, 10/IN.1/I3000 and 08/RFP/MTR/1659), and EU FP7 under the Marie Curie Reintegration Grant (PERG07-GA-2010-268300).

\newpage

\begin{figure}
\begin{center}
\includegraphics[width=\textwidth]{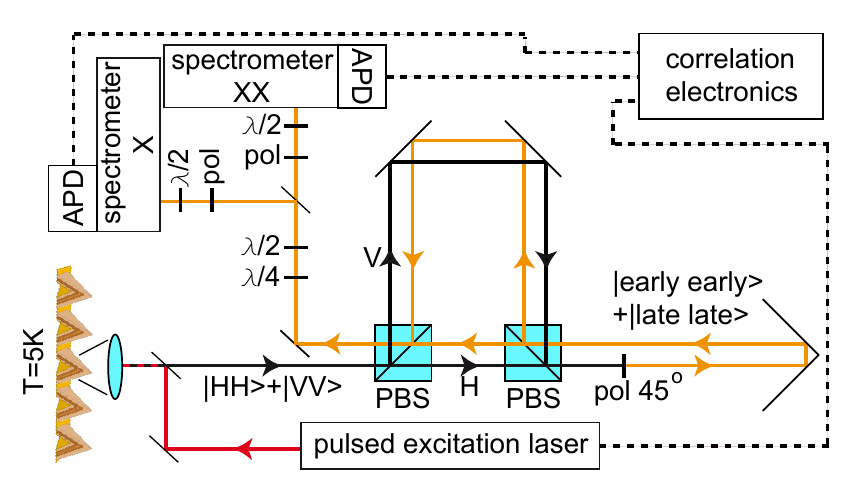}
\caption{Our setup, based on a quantum dot and a polarization-time-bin interface. Single polarization entangled photon pairs from the quantum dot are converted into single time-bin entangled photon pairs. A polarizer at $45^{\circ}$ erases all polarization entanglement. The time-bin measurement is performed via the orange path.}
\end{center}
\end{figure}

\begin{figure}
\begin{center}
\includegraphics[width=0.8\textwidth]{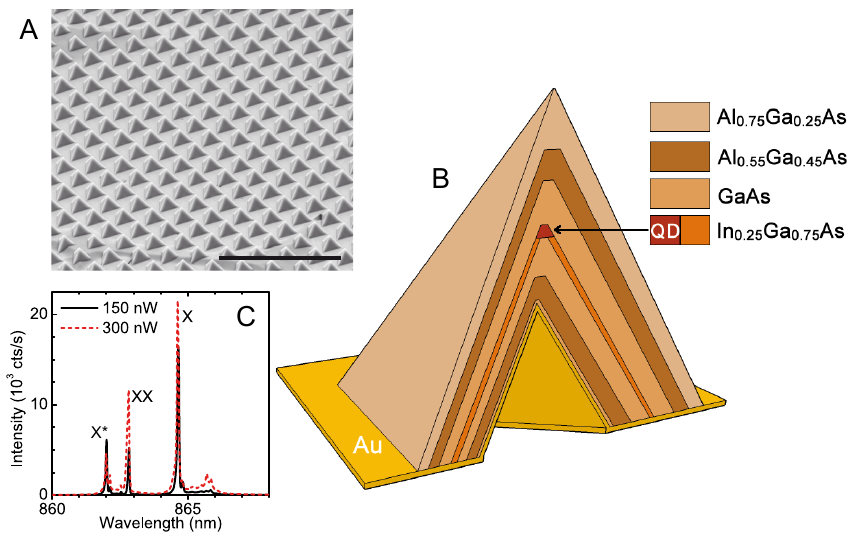}
\caption{Pyramidal quantum dot sample. (a) Scanning electron microscopy image of the sample, containing an array of pyramids with a 7.5 $\mu$m pitch. Each pyramid contains a quantum dot. Scale bar: 30 $\mu$m. Tilt angle: $60^{\circ}$. (b) Sketch of the internal epitaxial layer structure of a pyramid. The quantum dot is marked in red. (c) Emission spectrum of one quantum dot at two excitation powers. Indicated are the exciton (X), biexciton (XX) and trion (X*) emission lines.}
\end{center}
\end{figure}

\begin{figure}
\begin{center}
\includegraphics[width=\textwidth]{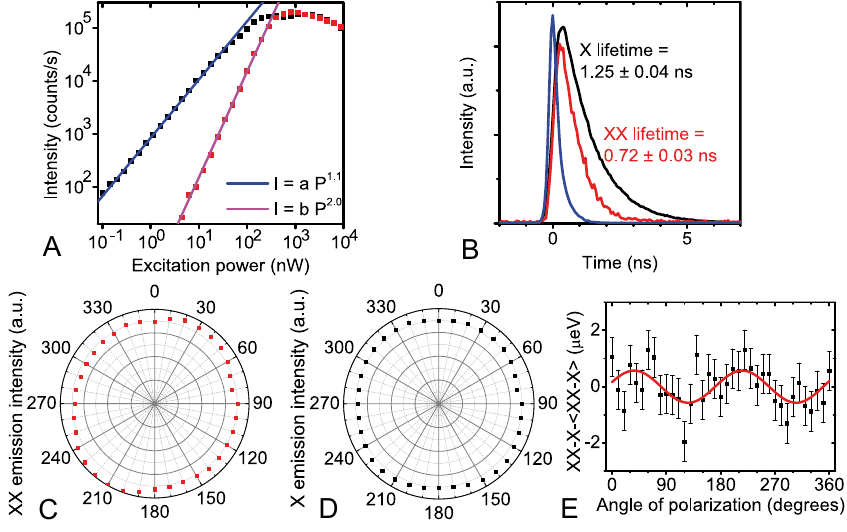}
\caption{Optical emission measurements on one pyramidal quantum dot. (a) The XX (red) and X (black) emission intensities versus excitation power. Fits to power functions show a quadratic dependence (power 2.0) for the XX emission and a nearly linear dependence (power 1.1) for the X emission. (b) Time-resolved photoluminescence measurements of the XX (red) and X (black) emission lines, revealing the lifetimes of both states. The blue curve is a measurement of the laser pulse and shows the time resolution of the detection system: the full width at half maximum is 0.34 ns. (c) XX and (d) X emission intensity versus polarization angle. (e) Energy separation between XX and X versus polarization angle. This measurement determines the fine-structure splitting: the energy splitting between the two X spin states is $0.6\pm0.2$ $\mu$eV.}
\end{center}
\end{figure}

\begin{figure}
\begin{center}
\includegraphics{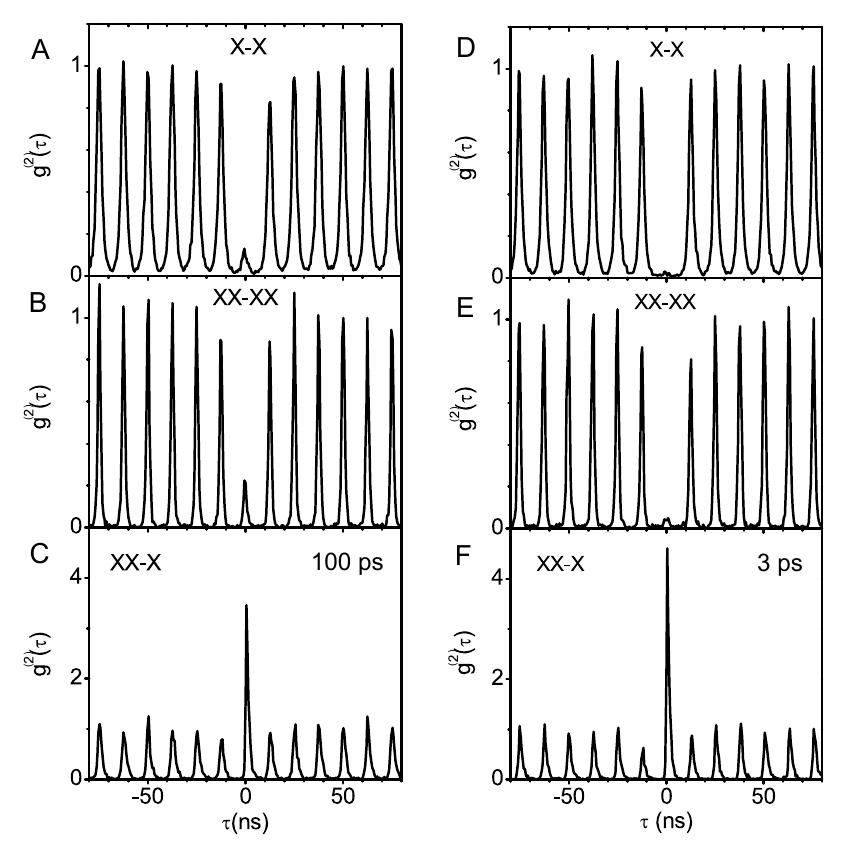}
\caption{Single-photon time-resolved correlation measurements. (a) X-X autocorrelation (start: X, stop: X), (b) XX-XX autocorrelation (start: XX, stop: XX), and (c) XX-X cross correlation (start: XX, stop: X) with 639-nm 100-ps 80-MHz excitation pulses at 150 nW. (d) X-X autocorrelation, (e) XX-XX autocorrelation, and (f) XX-X cross correlation with 750-nm 3-ps 80-MHz excitation pulses at 150 nW.}
\end{center}
\end{figure}

\begin{figure}
\begin{center}
\includegraphics[width=0.8\textwidth]{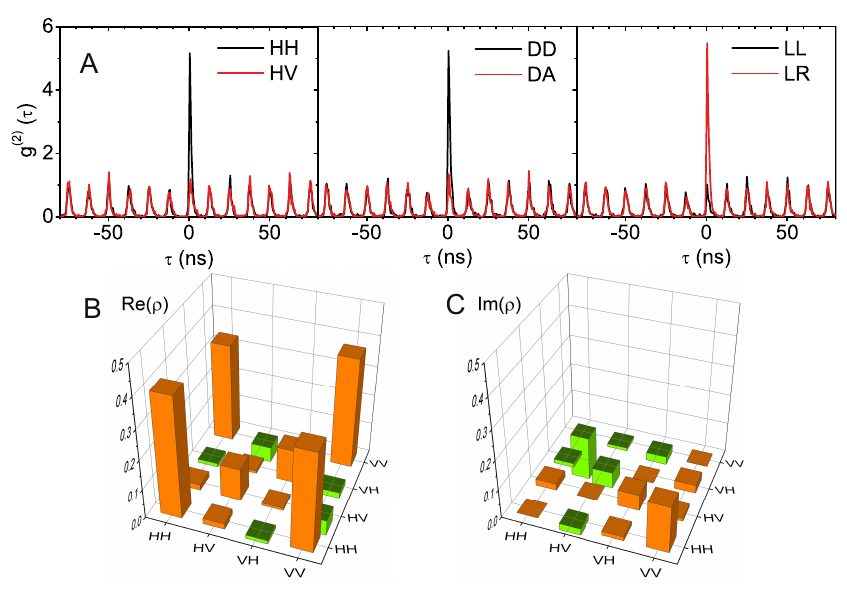}
\caption{(a) Measurements of correlations between the polarization of the XX photon and the polarization of the X photons. The first letter represents the polarization of the XX photon, the second letter the polarization of the X photon, where $H$, $V$, $D$, $A$, $L$, and $R$ stand for horizontal, vertical, diagonal, antidiagonal, left-handed, and right-handed polarization, respectively. (b) Real and (c) imaginary parts of the measured density matrix of the polarization quantum state of the two photons from the XX-X cascade. The positive matrix elements are orange, the negative green.}
\end{center}
\end{figure}

\begin{figure}
\begin{center}
\includegraphics[width=\textwidth]{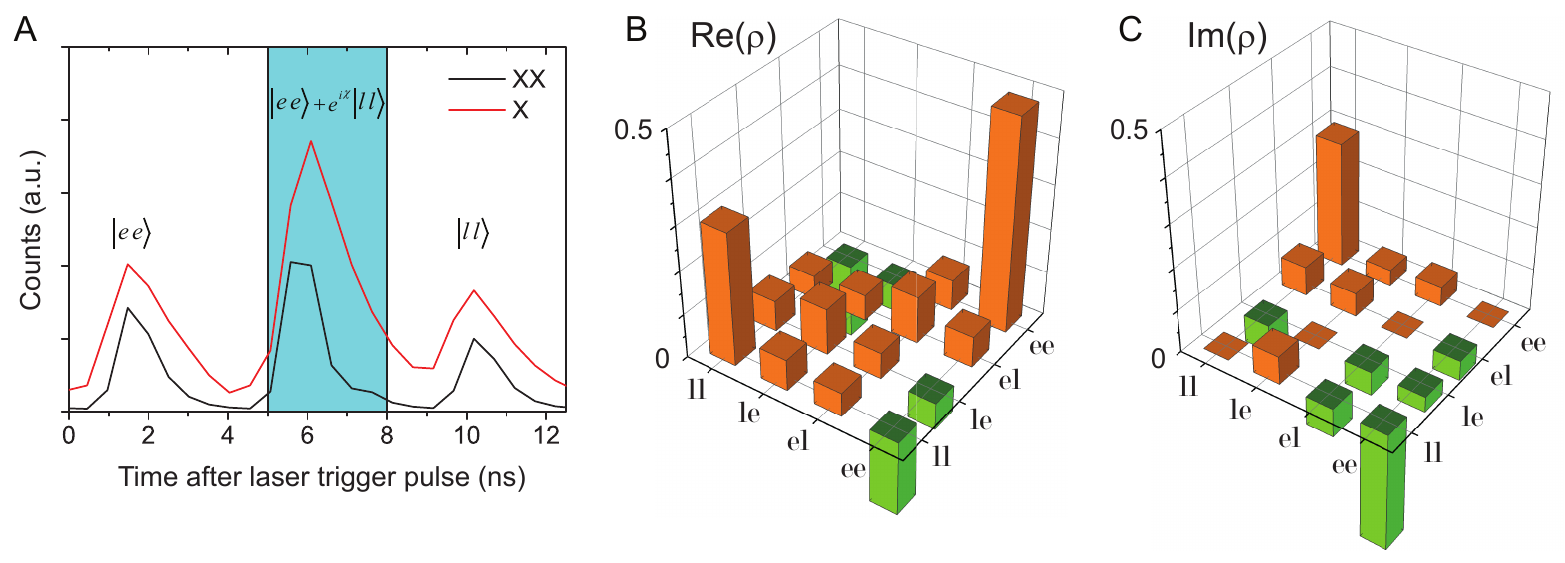}
\caption{Time-bin entanglement. (a) Count rate versus arrival time at the detector with respect to the laser trigger pulse. Post-selection on the photons in the blue time window. (b) Real and (c) imaginary part of the measured density matrix in the time-bin bases. The first and second letter stand for the time bin of the XX photon and the X photon, respectively.}
\end{center}
\end{figure}

\end{document}